\documentstyle[preprint,eqsecnum,aps,amsfonts]{revtex}

\newcommand{\be}{\begin{equation}}
\newcommand{\ee}{\end{equation}}
\newcommand{\bal}{\begin{array}{l}}
\newcommand{\eal}{\end{array}}
\newcommand{\bea}{\begin{eqnarray}}
\newcommand{\eea}{\end{eqnarray}}

\newcommand{\adot}{\dot{\alpha}}
\newcommand{\bdot}{\dot{\beta}}
\newcommand{\Bbar}{\bar{B}}
\newcommand{\BoldC}{\small{\Bbb C}}
\newcommand{\BoldH}{\small{\Bbb H}}
\newcommand{\BoldR}{\small{\Bbb R}}
\newcommand{\dete}{{\rm det}}
\newcommand{\e}{\epsilon}
\newcommand{\half}{\frac{1}{2}}
\newcommand{\Lambdab}{\Lambda^{b_0}}
\newcommand{\Lambdabi}{\Lambda_{i}^{b_{0i}}}
\newcommand{\Mhat}{M}
\newcommand{\Mtilde}{M}
\newcommand{\Nhat}{N}
\newcommand{\Ntilde}{N}
\newcommand{\Phat}{P}
\newcommand{\Ptilde}{P}
\newcommand{\NA}{N_A}
\newcommand{\NC}{N_C}
\newcommand{\NCL}{N_{CL}}
\newcommand{\NF}{N_F}
\newcommand{\NFL}{N_{FL}}
\newcommand{\Otwo}{O_2}
\newcommand{\Othree}{O_3}
\newcommand{\Otwofour}{O_2^{(4)}}
\newcommand{\Pf}{{\rm Pf}}
\newcommand{\Tr}{{\rm Tr}}
\newcommand{\Wdyn}{W_{\rm \scriptsize dyn}}

\def\footnoterule{\kern-5.25pt\hrule width1.5in\kern4.85pt}

\begin{document}


\baselineskip 20pt
\begin{titlepage}
 
\begin{flushright}
CALT-68-2062 \\
July 1996
\end{flushright}
 
\vskip 0.2truecm
 
\begin{center}
{\Large {\bf Symplectic SUSY Gauge Theories}}
 \vskip .05cm
{\Large {\bf with Antisymmetric Matter}}
\end{center}
 
\vskip 0.8cm
 
\begin{center}
Peter Cho and Per Kraus
\vskip .25 cm
{Lauritsen Laboratory \\
 California Institute of Technology \\
 Pasadena CA 91125}
\end{center}
 
\vskip 2.2cm
 
\begin{center}
{\bf Abstract}
\end{center}
 
        We investigate the confining phase vacua of supersymmetric $Sp(2\NC)$
gauge theories that contain matter in both fundamental and antisymmetric
representations.  The moduli spaces of such models with $\NF=3$ quark flavors
and $\NA=1$ antisymmetric field are analogous to that of SUSY QCD with
$\NF=\NC+1$ flavors.  In particular, the forms of their quantum superpotentials 
are fixed by classical constraints.  When mass terms are coupled to 
$W_{(\NF=3,\NA=1)}$ and heavy fields are integrated out, complete towers of
dynamically generated superpotentials for low energy theories with fewer
numbers of matter fields can be derived.  Following this approach, we deduce
exact superpotentials in $Sp(4)$ and $Sp(6)$ theories which cannot be
determined by symmetry considerations or integrating in techniques.  Building
upon these simple symplectic group results, we also examine the ground state
structures of several $Sp(4) \times Sp(4)$ and $Sp(6)~\times~Sp(2)$ models.  We 
emphasize that the top-down approach may be used to methodically find dynamical 
superpotentials in many other confining supersymmetric gauge theories. 
 
\end{titlepage}
 
\maketitle 

\newpage
 
\section{Introduction}

	During the past few years, dramatic progress has been made in 
understanding nonperturbative aspects of $N=1$ supersymmetric gauge theories.  
Recent work initiated by Seiberg and collaborators has addressed questions 
that previously seemed intractably difficult \cite{IntriligatorSeiberg}.
For example, the phase structures of several supersymmetric theories are now 
known, and dynamical mechanisms for various phase transitions have been 
explored \cite{IntriligatorSeibergII}.  Highly nontrivial exact superpotentials 
describing low energy limits have also been derived in many cases \cite{ILS}.  
Results from SUSY model investigations have shed light upon such interesting 
nonperturbative phenomena as confinement and chiral symmetry breaking.  
General insights gleaned from their study will hopefully be applicable to 
nonsupersymmetric gauge theories as well.

	Many of the new ideas about $N=1$ supersymmetric models have been 
developed within the context of SUSY QCD \cite{SeibergI,SeibergII}.
The low energy structure of this theory crucially depends upon its number of 
colors $\NC$ and quark flavors $\NF$.  For $\NF \le \NC+1$, SUSY 
QCD confines in the far infrared.  In the particular case when $\NF = \NC+1$, 
the quantum description of its moduli space of vacua is concisely summarized 
by the dynamically generated superpotential 
\be
\label{NFequalsNCplusone}
{W_{\scriptscriptstyle \NF=\NC+1} = 
{\Bbar M B - \dete M \over \Lambda^{2\NC-1}}.}
\ee
The meson and baryon superfields appearing in this expression represent gauge 
invariant coordinates on the moduli space.  The equations of motion obtained by 
varying $W$ with respect to $M$, $B$ and $\Bbar$ reproduce the constraints 
\begin{eqnarray}
M^i_j B^j = \Bbar_i M^i_j &=& 0 \nonumber\\
{1 \over \NC !} \e^{j_1 \cdots j_{\NF}} \e_{i_1 \cdots i_{\NF}} 
  M^{i_1}_{j_1} \cdots M^{i_{\NC}}_{j_{\NC}} &=& \Bbar_{i_{\NF}} B^{j_{\NF}}
\end{eqnarray}
which characterize the moduli space manifold.  It is important to note that 
the form of the superpotential in (\ref{NFequalsNCplusone}) can be deduced by 
requiring that these constraints be recovered.  As we shall see, this 
observation provides the key to unlocking the vacuum structure of many other 
confining supersymmetric gauge theories that are more complicated than SUSY 
QCD. 

	Once $W_{\NF=\NC+1}$ is known, it is straightforward to add tree level 
mass terms and flow down to theories with fewer numbers of quark flavors.  
After integrating out the first quark, one finds that no dynamically generated 
superpotential exists in $\NF=\NC$ SUSY QCD.  Nonetheless, nonperturbative 
effects do transmute the classical moduli space relation $\dete M - B\Bbar=0$ 
into the quantum constraint \cite{SeibergI}
\be
\label{SQCDquantumconstraint}
{\dete M - B\Bbar = \Lambda^{2\NC}.}
\ee
When additional quark fields are integrated out, the superpotential for the 
resulting low energy effective field theory with $\NF < \NC$ flavors takes 
the form
\be
\label{NFlessNC}
{W_{\scriptscriptstyle \NF < \NC} = 
(\NC - \NF) \Bigl[ {\Lambda^{3\NC - \NF} \over \dete M } 
\Bigr]^{1/(\NC-\NF)} .}
\ee
The vacua in this case may be stabilized by adding mass terms to the 
dynamically generated superpotential.

	The ground state structures of other confining supersymmetric gauge 
theories with matter in only defining representations are qualitatively 
similar to that for SUSY QCD \cite{IntriligatorSeibergIII,IntriligatorPouliot}.
Symmetry and holomorphy considerations fix the functional forms of the 
low energy superpotentials in such 
models.  On the other hand, the quantum moduli spaces of theories with more 
complicated matter content are generally much harder to uncover.  Dynamical 
superpotentials for $SU(\NC)$ gauge theories with $\NF \le 3$ fundamentals, 
$\NC+\NF-4$ antifundamentals and one antisymmetric tensor were derived in 
ref.~\cite{Poppitz}.  The vacuum structure for some confining $SO(\NC)$ models 
with matter in both vector and spinor representations have also been studied 
\cite{Murayama,Kawano}.  But the list of solved multimatter theories is not 
very long.

	In this article, we will analyze the moduli spaces for a class of 
symplectic models that contain matter in both fundamental and antisymmetric 
representations.  We focus upon the low energy descriptions of their confining 
phase sectors.
\footnote{$\>$ Other phases of such theories, in the presence of tree 
level superpotentials, have been studied in ref.~\cite{IntriligatorI}.}
The methods we develop to construct the superpotentials for these symplectic 
theories can be applied to other simple group models.  They also expand the 
number of known product group moduli spaces.  These double-matter models 
consequently provide useful laboratories for exploring new aspects of 
supersymmetric gauge theories. 

	Our paper is organized as follows.  We first review some basic 
elements of symplectic group theory in section~2.  We then study the 
vacuum structure of $Sp(2\NC)$ models which contain antisymmetric 
matter in section~3.  We discuss in detail the superpotentials for $Sp(4)$ 
and $Sp(6)$ theories with $\NF \le 3$ quark flavors and one antisymmetric 
field.  In section~4, we use these simple group superpotentials to map the 
moduli spaces of several $Sp(4) \times Sp(4)$ and $Sp(6) \times Sp(2)$ product 
group models.  Finally, we close in section~5 with a summary of our findings 
and some thoughts on possible future extensions of this work. 

\section{Sp(2N) basics}

	The definition of the classical group of symplectic transformations 
stems from geometrical considerations similar to those for the more 
familiar orthogonal and unitary groups.  The fundamental representations 
of $SO(N)$, $SU(N)$ and $Sp(2N)$ leave invariant the inner products 
\begin{mathletters}
\begin{eqnarray}
\label{innerproducts}
\langle {\bf v}_1, {\bf v}_2 \rangle &= &
	{\bf v}_1^T \cdot {\bf v}_2  \label{innerproducts-a} \\
\langle {\bf z}_1, {\bf z}_2 \rangle  &= &
	{\bf z}_1^\dagger \cdot {\bf z}_2  \label{innerproducts-b} \\
\langle {\bf q}_1, {\bf q}_2 \rangle &= &
	\bar{\bf q}_1^T \cdot {\bf q}_2 \label{innerproducts-c} 
\end{eqnarray}
\end{mathletters}
\noindent
defined on the real, complex and quaternionic vector spaces ${\BoldR}^N$, 
${\BoldC}^N$ and ${\BoldH}^N$.  $Sp(2N)$ is thus isomorphic to the group of 
$N \times N$ matrices with quaternionic elements which preserve the dotproduct 
in (\ref{innerproducts-c}).    

	Just as any complex number may be regarded as an ordered pair of two 
real numbers, so may any quaternion be viewed as an ordered pair of two 
complex numbers.  An element $q \in \BoldH$ decomposes over ${\BoldC}^2$ and 
${\BoldR}^4$ as 
\begin{eqnarray}
q &=& z_1 + j z_2 \qquad\qquad\qquad\qquad\quad z_1, z_2 \in {\BoldC}\nonumber\\
 & =& (\alpha + i \beta) + j (\gamma - i \delta) 
  \qquad\> \alpha,\beta,\gamma,\delta \in {\BoldR} \\
  & =& \alpha + i \beta + j \gamma + k \delta \nonumber
\end{eqnarray}
where the symbols $i$, $j$ and $k$ satisfy the relations $i^2=j^2=k^2=-1$
and $ij=-ji=k$ plus cyclic permutations.  Quaternionic vectors in ${\BoldH}^N$
can similarly be rewritten as elements of ${\BoldC}^{2N}$: 
\begin{eqnarray}
{\bf q}_1 & \equiv &{\bf s} + j {\bf t} \nonumber \\
{\bf q}_2 & \equiv &{\bf u} + j {\bf v}.
\end{eqnarray}
Recalling that $i$, $j$ and $k$ are mapped into their negatives under 
conjugation, we see that the 
\vfill\eject\noindent
inner product (\ref{innerproducts-c}) is expressible as 
\be
\label{qinnerproduct}
{\langle {\bf q}_1, {\bf q}_2 \rangle = 
\pmatrix{ s_1^* & t_1^* & \cdots & s_N^* & t_N^* \cr } 1_{2N \times 2N} 
 \pmatrix{ u_1 \cr
	  v_1 \cr
	  \vdots \cr
	  u_N \cr
	  v_N \cr} + 
j \pmatrix{ s_1^T & t_1^T & \cdots & s_N^T & t_N^T \cr } J 
\pmatrix{ u_1 \cr
	  v_1 \cr
	  \vdots \cr
	  u_N \cr
	  v_N \cr}} 
\ee
where $J = 1_{N \times N} \times i \sigma_2$.  Any rotation $U$ satisfying 
$U^\dagger U = 1$ and $U^T J U = J$ preserves the RHS of eqn.~(\ref{qinnerproduct}).  
The set of all such complex $2N \times 2N$ matrices forms the fundamental 
irreducible representation of $Sp(2N)$.  

	$Sp(2N)$ is a rank $N$ subgroup of $U(2N)$ with $2N^2 + N$ generators.  
One convenient basis for the symplectic group's generators in the $2N$ 
dimensional representation is schematically given by the tensor products 
${\vec S}_{N \times N} \times {\vec \sigma}$ and 
$A_{N \times N} \times 1_{2 \times 2}$ where $S$ and $A$ respectively denote 
symmetric and antisymmetric hermitian generators of $SU(N)$ \cite{Georgi}. 
In this basis, one can readily check that the similarity transformation 
$-T_a^* = J T_a J^{-1}$ holds for all $2N^2+N$ generators of the $2N$ and 
$\overline{2N}$ representations.  The fundamental irrep of $Sp(2N)$ is 
consequently pseudoreal.   All other representations formed by taking tensor 
products of fundamentals are also either real or pseudoreal.  No essential 
distinction therefore exists between matter and antimatter in symplectic gauge 
theories.  

	Although $Sp(2N)$ may seem more foreign than $SO(N)$ and $SU(N)$, 
symplectic group theory is actually easier than its orthogonal and unitary 
analogs.  This fact simplifies the analysis of the symplectic models that we 
shall study in this article. 

\section{Symplectic models with antisymmetric matter}

	The number of different classes of supersymmetric gauge theories 
which possess nontrivial confining phases is surprisingly small.  Such models 
must first be asymptotically free.  As is well known, only a limited 
number of theories with matter in relatively low dimension representations 
exhibit asymptotic freedom.  Additional conditions that confining models must 
satisfy significantly restrict their number.  As a result, only a handful of 
confining simple group classes exist.  

	In order to determine whether a SUSY model confines, it is 
useful to consider the R-charge associated with its strong interaction 
scale.  Recall that any $N=1$ supersymmetric gauge theory with zero tree 
level superpotential can be rendered invariant under a $U(1)_R$ symmetry which 
rotates the Grassmann $\theta$ parameter by a phase.  By definition, $\theta$ 
has one unit of R-charge.  The overall R-charges of matter superfields are 
{\it a~priori} unspecified.  If they are all set to zero, the strong 
interaction scale must be assigned 
\be
\label{Rcharge}
{R \bigl( \Lambdab \bigr) = K(\rm Adj) - 
\sum_{\rm{\buildrel matter \over {\scriptscriptstyle reps \>\rho}}} K(\rho)}
\ee
to cancel a global $U(1)_R$ anomaly in the quantized theory.  In this 
expression, $K(\rho)$ denotes the group theory index of representation $\rho$ 
with $K({\rm fundamental}) \equiv 1$.  This last normalization 
choice fixes the one-loop beta function coefficient
\be
\label{bzero}
{b_0 = \half \bigl[ 3 K({\rm Adj}) - 
\sum_{\rm{\buildrel matter \over {\scriptscriptstyle reps \>\rho}}} K(\rho)
\bigr]}
\ee
that appears on the LHS of (\ref{Rcharge}). 

	Any dynamically generated superpotential $\Wdyn$ must have $R=2$ in 
order for the supersymmetric action to be $U(1)_R$ invariant.  The strong 
interaction scale dependence of $\Wdyn$ is consequently determined in simple 
group models since only $\Lambdab$ carries nonvanishing R-charge.  In SUSY 
QCD, $\Lambdab$ enters into the denominator of superpotential 
(\ref{NFequalsNCplusone}) when $R(\Lambdab)=2(\NC-\NF) = -2$.  The numerator 
is then a simple polynomial in the meson and baryon fields.  If one attempts 
to construct $\Wdyn$ for $\NF=\NC+2$ flavors, a square root branch point is 
encountered at the moduli space origin.  Such a singularity indicates that 
new phenomena emerge in $\NF=\NC+2$ SUSY QCD which are absent in the 
$\NF=\NC+1$ theory.  Indeed, it is now known that supersymmetric QCD ceases 
to confine at this juncture and enters into the free magnetic phase 
\cite{SeibergII}.  SUSY QCD therefore binds together colored partons into 
colorless hadrons only so long as $R(\Lambdab) \ge -2$.  

	Similar heuristic arguments suggest that adding sufficient matter 
into any SUSY model causes its would-be superpotential to develop a branch 
point at the origin of moduli space which signals the end of the confining 
phase.  If this hypothesis is accepted, it is straightforward to check that 
the number of confining model classes is quite limited. 

	One interesting set of theories which does possess a nontrivial 
confining phase is based upon the symmetry group 		
\be
\label{symgroup}
{G = Sp(2\NC)_{\rm local} \times \left[ SU(2\NF) \times U(1)_Q
\times U(1)_A \times U(1)_R \right]_{\rm global}}
\ee
with microscopic matter 
\begin{eqnarray}
Q & \sim & (2 \NC; 2 \NF; 1,0,0)\nonumber \\
A & \sim & \Bigl[ {2 \NC \choose  2} - 1; 1; 0,1,0 \Bigr]  \\
\Lambda^{b_0} & \sim & \bigl(1; 1; 2\NF, 2 (\NC-1), 4 - 2\NF \bigr). \nonumber
\end{eqnarray}
Several points about the field content of these symplectic theories should be 
noted.  Firstly, in order to avoid a global Witten anomaly, these models must 
involve an even number $2 \NF$ of quarks in the fundamental irrep of $Sp(2\NC)$ 
\cite{Witten}.  In the limit of zero tree level superpotential, the 
supersymmetric action remains invariant under a global $U(2\NF) \simeq 
SU(2\NF) \times U(1)_Q$ symmetry that rotates the quarks among themselves.  
Secondly, the $A$ field transforms according to the ``traceless'' antisymmetric 
representation of $Sp(2\NC)$.  Its inner product with the skew metric $J$ 
vanishes.  Finally, we regard the strong interaction scale $\Lambda^{b_0}$ with 
$b_0 = 2\NC -\NF + 4$ as a background ``spurion'' field \cite{SeibergIII}.
Its abelian charges are chosen so that all global $U(1)$ factors in $G$ 
are nonanomalous.  Looking at the R-charge assignment for $\Lambda^{b_0}$, we 
see that these models confine so long as their number of quark flavors does 
not exceed $\NF=3$. 

	The classical moduli space of vacua for the symplectic theories is 
simplest to analyze when $\NF=0$.  The flat directions along which the scalar 
potential vanishes are then determined by the antisymmetric field expectation 
values satisfying $\Tr  (T_a A A^\dagger) = 0$.  Working with the basis of 
fundamental irrep $Sp(2\NC)$ generators introduced in section~2, we find that 
the general solution to this D-flatness condition is given by a linear 
combination of $U(2\NC)/Sp(2\NC)$ coset space generators:
\be
\label{Dflatnesssoln}
{A A^\dagger = S_{\NC \times \NC} \times 1_{2 \times 2}
+ {\vec A}_{\NC \times \NC} \times {\vec \sigma}.}
\ee
Since the quaternions $(1,i,j,k)$ are isomorphic to the $2\times 2$ 
Pauli matrices $(1_{2 \times 2},-i {\vec\sigma})$, the matrix $A A^\dagger$ 
may be regarded as a general hermitian element of ${\BoldH}^{\NC \times \NC}$.  
It can be diagonalized by some unitary matrix built out of quaternions that
is equivalent to an element of $Sp(2\NC)$:
\footnote{$\>$ We thank Howard Georgi for this group theory insight.}
\be
\label{diagsoln}
{A A^\dagger \to D_{\NC \times \NC} \times 1_{2 \times 2}.}
\ee
The vev of $A$ itself looks like
\be
\label{Avev}
{\langle A \rangle = 
\pmatrix{ z_1 &        & \cr
  	      & \ddots & \cr
	      &        & z_{\NC}} \times i \sigma_2 
\qquad {\rm with} \qquad
\sum_{n=1}^{\NC} z_n = 0}
\ee
up to a local gauge transformation.  The classical moduli space is thus 
$\NC-1$ complex dimensional.

	The flat directions in the $\NF=0$ theory are labeled by the gauge 
invariant operators
\be
\label{Oops}
{O_n = \Tr\bigl[ (AJ)^n \bigr], \qquad n=2,3,\cdots,\NC.}
\ee
In the models with $\NF=1$, 2 and 3 quark flavors, additional meson fields
\begin{eqnarray}
\label{mesonops}
M_{ij} &=& Q_i^T J Q_j \nonumber \\*
N_{ij} &=& Q_i^T J A J Q_j \nonumber \\*
P_{ij} &=& Q_i^T J A J A J Q_j \\*
& \vdots & \nonumber \\*
R_{ij} &=& Q_i^T J (AJ)^{\NC-1} Q_j \nonumber
\end{eqnarray}
are needed to act as moduli space coordinates.  The characteristic polynomial 
for the matrix $AJ$ truncates their number.  Only $\NC$ such meson operators 
are therefore independent for an $Sp(2\NC)$ color group.  
	
	The simplest symplectic models that support antisymmetric matter 
have gauge group $Sp(4)$.  The vacuum structure of these theories 
is most readily analyzed if we start with $\NF=3$ quark flavors.  The 
hadron superfield charge assignments for this case are listed in Table~1.  
Looking at the $R=-2$ entry for the spurion field $\Lambdab$, we see that the 
$\NF=3$, $\NA=1$ $Sp(4)$ model is analogous to $\NF=\NC+1$ SUSY QCD.  Its
classical and quantum moduli spaces are the same, and the constraint equations 
that define its moduli space manifold are polynomials in the gauge invariant 
hadron fields.  Symmetry considerations significantly limit the possible terms 
in the dynamical superpotential 
\be
\label{NFequalthree}
{\Wdyn = { {\rm Some \ polynomial \ in \ } M_{ij}, 
N_{ij} {\rm \ and \ } O_2 \over \Lambda^5}}
\ee
which generates these constraints.  In order for $\Wdyn$ to be 
$U(1)_Q \times U(1)_A$ invariant, the numerator in eqn.~(\ref{NFequalthree}) 
must have abelian charges $Q=6$ and $A=2$.  Moreover, the numerator's 

\vfill\eject

\begin{centering}
\begin{tabular}{|c|cccc|}  \hline
$\quad$ Superfield $\quad$ & $\qquad U(1)_Q \qquad$ & $\qquad U(1)_A \qquad$ 
	   		   & $\qquad U(1)_R \qquad$ & Mass Dimension
\rule[-.25cm]{0cm}{.7cm} \\ \hline
$M_{ij}$ & $2$ & $0$ & $0$ & $2$ \rule[-.25cm]{0cm}{.7cm}  \\
$N_{ij}$ &  2 & 1 & 0 & 3 \rule[-.25cm]{0cm}{.7cm} \\
$O_2$ &  0 & 2 & 0 & 2 \rule[-.25cm]{0cm}{.7cm} \\
$\Lambda^5_{(3,1)}$ & 6 & 2 & -2 & 5 \rule[-.25cm]{0cm}{.7cm} \\ \hline
\end{tabular}

\bigskip
\centerline{Table 1: $U(1)$ charge assignments in the $\NF=3$, $\NA=1$ $Sp(4)$ 
 theory}

\end{centering}

\bigskip\bigskip\noindent
mass dimension must equal 8 so that $\Wdyn$ has dimension 3.  The form 
of the superpotential consistent with all these restrictions is unique:
\be
\label{NFequalthreeII}
{\Wdyn = - {\e^{ijklmn} \over 48 \Lambda^5} 
\bigl[ M_{ij} M_{kl} M_{mn} O_2 + \alpha N_{ij} N_{kl} M_{mn} \bigr].}
\ee

	Symmetry principles leave undetermined the overall numerical prefactor 
that multiplies $\Wdyn$.  Its value is rather arbitrary since it may be 
altered by a redefinition of the strong interaction scale.  For 
later convenience, we have set this normalization factor equal to 
$-1/48$.  On the other hand, the value of the relative coefficient $\alpha$ 
between the two terms in (\ref{NFequalthreeII}) is fixed by consistency 
requirements.  The equations of motion obtained by varying $\Wdyn$ with 
respect to $M_{ij}$, $N_{ij}$ and $O_2$ yield the classical constraints
\begin{mathletters}
\begin{eqnarray}
\e^{ijklmn} \bigl[ 3 O_2 M_{kl} M_{mn} + \alpha N_{kl} N_{mn} \bigr] &=& 0 
 \label{constraints-a} \\
\e^{ijklmn} N_{kl} M_{mn} &=& 0  \label{constraints-b} \\
\e^{ijklmn} M_{ij} M_{kl} M_{mn} &=& 0  \label{constraints-c} 
\end{eqnarray}
\end{mathletters}
\noindent
which must be satisfied when the hadron fields are decomposed in terms 
of their parton constituents.  Inserting $M_{ij} = Q^T_i J Q_j$, 
$N_{ij} = Q^T_i J A J Q_j$ and $O_2 = \Tr\bigl[( AJ)^2 \bigr]$ into these 
relations, we find (\ref{constraints-b}) and 
(\ref{constraints-c}) are classical identities whereas 
(\ref{constraints-a}) is satisfied only if $\alpha=12$.  

	In order to probe the structure of the $\NF=3, \NA=1$ $Sp(4)$ theory's 
moduli space, it is instructive to couple tree level sources to the dynamical 
superpotential in (\ref{NFequalthreeII}).  We choose to add the $Q$ and $A$ 
field mass terms
\be
\label{Wtree}
{W_{\rm tree} = \half \mu^{ij} M_{ji} + m O_2}
\ee
which lift all flat directions while preserving the gauge group.  The 
hadron operators then develop the expectation values 
\begin{eqnarray}
\label{vevs}
\langle M_{ij} \rangle &=& m^{\frac{1}{3}} (\Pf\mu)^{\frac{1}{3}} \Lambda^{\frac{5}{3}} 
(\mu^{-1})_{ij} \nonumber \\
\langle N_{ij} \rangle &=& 0 \\
\langle O_2 \rangle &=& {1 \over 6} m^{-\frac{2}{3}}  (\Pf\mu)^{\frac{1}{3}}
\Lambda^{\frac{5}{3}} \nonumber 
\end{eqnarray}
in the presence of the sources.  If we substitute these vevs back into the 
superpotential, we can methodically integrate out quark flavors and flow down 
to $Sp(4)$ theories with smaller values of $\NF$.  The tower of resulting 
dynamical superpotentials is displayed below:
\bigskip
\begin{eqnarray} \label{spfourflows} 
& W_{(\NF=3,\NA=1)} = -\displaystyle{\e^{ijklmn} \over 48 \Lambda_{(3,1)}^5} 
\bigl[ M_{ij} M_{kl} M_{mn} \Otwo + 12 N_{ij} N_{kl} M_{mn} \bigr] &
\nonumber\\*
& \downarrow & \nonumber\\*
& W_{(\NF=2,\NA=1)} = X \e^{ijkl} \bigl[ M_{ij} M_{kl} \Otwo + 
4 N_{ij} N_{kl} - 8 \Lambda_{(2,1)}^6 \bigr] + Y \e^{ijkl} M_{ij} N_{kl} &
\nonumber\\*
& \downarrow & \nonumber\\*
& W_{(\NF=1,\NA=1)} = \displaystyle {\Pf M \>\> \Lambda_{(1,1)}^7 \over 
\Otwo (\Pf M)^2 - 4 (\Pf N)^2} & \nonumber\\*
& \swarrow \qquad\qquad\qquad\qquad\qquad \searrow & \nonumber\\*
& W_{(\NF=0,\NA=1)} = 2 \Bigl[ \displaystyle{\Lambda_{(0,1)}^8 \over \Otwo} 
\Bigr]^{\half} \qquad\qquad\qquad W_{(\NF=0,\NA=1)}=0 & \nonumber\\*
& \downarrow \qquad\qquad\qquad\qquad\qquad\qquad & \nonumber\\*
& W_{(\NF=0,\NA=0)} = 3 \bigl[ \Lambda_{(0,0)}^9 \bigr]^{\frac{1}{3}}. 
\qquad\qquad\qquad\qquad\qquad\qquad\qquad &
\end{eqnarray}

	All the strong interaction scales appearing in this hierarchy are 
related by matching conditions that ensure continuity of the running $Sp(4)$ 
gauge coupling across heavy particle thresholds.  These matching relations are 
most conveniently expressed in the basis where the quark mass matrix 
\be
\label{mumatrix}
{\mu = \pmatrix{\mu_1 & & \cr
			     & \mu_2 & \cr
			     & & \mu_3 \cr} \times i \sigma_2}
\ee
is block diagonalized.  The matching conditions then form the simple chain 
\be
\label{matching}
{\Lambda_{(0,0)}^9 = m \Lambda_{(0,1)}^8 = m \mu_1 
\Lambda_{(1,1)}^7 = m \mu_1 \mu_2 \Lambda_{(2,1)}^6 = 
m \mu_1 \mu_2 \mu_3 \Lambda_{(3,1)}^6.}
\ee
 
	A number of checks on the renormalization group flow between the 
different $Sp(4)$ theories in eqn.~(\ref{spfourflows}) can be performed.  Firstly, 
we can at any stage integrate out the antisymmetric $A$ field and recover the 
superpotential
\be
\label{Wquark}
{W_{(\NF,\NA=0)} = (3-\NF) 
\Biggl[ {\Lambda_{(\NF,0)}^{9-\NF} \over \Pf M } \Biggr]^{1/(3-\NF)} }
\ee
for $Sp(4)$ theory with $2\NF$ flavors of quarks \cite{IntriligatorPouliot}.
\footnote{$\>$ Our normalization for $\Lambdab$ in the pure quark theory 
differs by a factor of $2^{\NC-1}$ from the one adopted in 
ref.~\cite{IntriligatorPouliot}}
We used the normalization of this known result along with the strong 
interaction scale matching relations to fix the overall numerical prefactor 
in eqn.~(\ref{NFequalthreeII}).  Secondly, we can verify that the quantum 
constraints
\begin{eqnarray}
\label{spfourconstraints}
\e^{ijkl} \Bigl[ \Mtilde_{ij} \Mtilde_{kl} O_2 + 4 \Ntilde_{ij} \Ntilde_{kl} 
\Bigr] &=& 8 \Lambda_{(2,1)}^6 \nonumber \\
\e^{ijkl} \Mtilde_{ij} \Ntilde_{kl} &=& 0
\end{eqnarray}
obtained by varying $W_{(\NF=2,\NA=1)}$ with respect to its $X$ and $Y$ 
Lagrange multiplier superfields yield valid classical relations in the 
$\Lambda_{(2,1)} \to 0$ limit.  $\NF=2$, $\NA=1$ $Sp(4)$ theory is analogous 
to $\NF=\NC$ SUSY QCD inasmuch as $R(\Lambdab)=0$ in both models.  But whereas 
the moduli space of the latter is defined by just the single constraint in 
eqn.~(\ref{SQCDquantumconstraint}), the moduli space of the former 
involves the two relations in (\ref{spfourconstraints}).  In general, $\NC$ 
different equations are needed to characterize the $\NF=2$, $\NA=1$ 
$Sp(2\NC)$ moduli manifold.  Finally, the two $W_{(\NF=0,\NA=1)}$ 
superpotentials for $Sp(4)$ theory with matter in only the 5-dimensional 
antisymmetric irrep must agree with those for $SO(5)$ theory with matter in 
the vector representation since the Lie algebras of $Sp(4)$ and $SO(5)$ are 
identical.  A quick check confirms that they do \cite{ILS}.

	The basic strategy we followed to construct the tower of 
low energy $Sp(4)$ models can be applied to other symplectic theories in 
the same class with greater numbers of colors.  This procedure generally yields 
highly nontrivial superpotentials for complicated quantum moduli spaces.  Yet 
its starting point depends only upon classical physics.  The structure of the 
$\NF=3$, $\NA=1$ superpotential in $Sp(2\NC)$ theory is
\be
\label{genNFequalthree}
{W_{(\NF=3,\NA=1)} = 
{P(M_{ij}, N_{ij}, \cdots; O_2, O_3, \cdots) \over \Lambdab}}
\ee
where $P$ denotes some holomorphic function of the color-singlet superfields.  
Symmetry considerations do not prevent $P$ from being riddled with poles 
and branch cuts.  But such singularities would possess no clear physical 
interpretation.  Moreover, duality arguments suggest that a magnetic dual to 
the $\NF=3$, $\NA=1$ $Sp(2\NC)$ theory is very weakly coupled in the far 
infrared and does not generate any superpotential singularities 
\cite{SeibergII}.  As a result, it is reasonable to assume that $P$ is 
a polynomial.  The equations of motion obtained by varying $P$ with respect to 
its arguments then yield classical constraints which must be satisfied when 
the hadron operators are decomposed in terms of their underlying $Q$ and $A$ 
constituents.  This requirement fixes polynomial $P$.  The superpotentials for 
other models with smaller $\NF$ or $\NA$ values can subsequently be obtained 
from (\ref{genNFequalthree}) by systematically integrating out matter degrees 
of freedom.  

	To illustrate the utility of this approach, we consider $Sp(6)$ 
theory with $\NF=0$ and $\NA=1$.  We recall that the scalar potential in 
this model has two independent flat directions which are labeled by the gauge 
invariant operators $O_2 = \Tr \bigl[ (AJ)^2 \bigr]$ and 
$O_3 = \Tr \bigl[ (AJ)^3 \bigr]$.  The dimensionless and chargeless ratio 
$R~\equiv~-~12~O_3^2~/~O_2^3$ can be constructed from these two operators.  
Symmetry places no restrictions on the superpotential's dependence upon $R$.  
Furthermore, ``integrating in'' techniques fail to determine the functional 
form of $W_{(\NF=0,\NA=1)}$ \cite{ILS,IntriligatorII}.  But if we start with 
the $\NF=3$, $\NA=1$ $Sp(6)$ model and progressively integrate out quark 
flavors, we can in fact deduce the dynamical superpotential in the pure 
antisymmetric theory.  The tower of confining phase $Sp(6)$ superpotentials is 
displayed below:
\bigskip
\begin{eqnarray}\label{spsixflows}
& W_{(\NF=3,\NA=1)} = \displaystyle{\e^{ijklmn} \over 
48 \Lambda_{(3,1)}^7} \bigl[ 12 M_{ij} M_{kl} P_{mn} \Otwo 
- M_{ij} M_{kl} M_{mn} \Otwo^2 - 8 M_{ij} M_{kl} N_{mn} \Othree & \nonumber\\*
& \qquad\qquad\qquad
 - 48 N_{ij} N_{kl} P_{mn} - 48 P_{ij} P_{kl} M_{mn} \bigr] & \nonumber\\*
& \downarrow & \nonumber\\*
& W_{(\NF=2,\NA=1)} = X \e^{ijkl} \bigl[ 3 \Mtilde_{ij} \Mtilde_{kl} \Otwo^2 + 
  16 \Mtilde_{ij} \Ntilde_{kl} \Othree - 24 \Mtilde_{ij} \Ptilde_{kl} \Otwo + 
  48 \Ptilde_{ij} \Ptilde_{kl} - 24 \Lambda_{(2,1)}^8 \bigr]  \nonumber\\*
& \qquad +Y \e^{ijkl} \bigl[ \Mtilde_{ij} \Mtilde_{kl} \Othree + 12 
 \Ntilde_{ij} \Ptilde_{kl} \bigr] 
+ Z \e^{ijkl} \bigl[ \Mtilde_{ij} \Mtilde_{kl} \Otwo 
 - 4\Ntilde_{ij} \Ntilde_{kl} - 8\Mtilde_{ij} \Ptilde_{kl} \bigr] & \nonumber\\*
& \downarrow & \nonumber\\*
& W_{(\NF=1,\NA=1)} = 9 \bigl[ (\Pf \Mhat)(\Pf \Phat) - (\Pf \Nhat)^2 
\bigr] \>\> \Lambda_{(1,1)}^9  
\Bigl\{ 4 \Othree^2 (\Pf\Mhat)^3 + 12 \Otwo \Othree (\Pf\Mhat)^2 (\Pf\Nhat) &
\nonumber\\*
& \mbox{}\qquad + 9 {\Otwo}^2 (\Pf\Mhat)(\Pf\Nhat)^2 + 24 \Othree \bigl[ 
(\Pf\Nhat)^3 - 3 (\Pf\Mhat)(\Pf\Nhat)(\Pf\Phat)\bigr] \nonumber\\*
& \mbox{}-36 O_2 \bigl[ (\Pf\Nhat)^2 (\Pf\Phat)+ (\Pf\Mhat)(\Pf\Phat)^2 
\bigr] + 144 (\Pf\Phat)^3 \Bigr\}^{-1} & \nonumber\\*
& \qquad\qquad\qquad\qquad \swarrow 
\qquad\qquad\qquad\qquad\qquad\qquad\qquad \searrow & \nonumber\\*
& W_{(\NF=0,\NA=1)} = \displaystyle{4 \Big[ \Lambda_{(0,1)}^{10} \Big]^{\half} 
\over \Otwo \Big[ \bigl( \sqrt{R} + \sqrt{R+1} \bigr)^{\frac{2}{3}} 
+ \bigl( \sqrt{R} + \sqrt{R+1} \bigr)^{-\frac{2}{3}} -1 \Big] } 
\qquad W_{(\NF=0,\NA=1)}=0 & \nonumber\\*
& \downarrow \qquad\qquad\qquad\qquad\qquad\qquad & \nonumber\\*
& W_{(\NF=0,\NA=0)} = 4 \bigl[ \Lambda_{(0,0)}^{12} \bigr]^{\frac{1}{4}}. 
\qquad\qquad\qquad\qquad\qquad\qquad &
\end{eqnarray}

	The low energy $Sp(6)$ effective theories satisfy several consistency 
checks.  Known $Sp(6)$ models with just quark matter are recovered when the 
antisymmetric field is integrated out.  't~Hooft anomaly matching conditions 
are also satisfied at various points on the moduli manifolds.  For instance, 
the global $SU(4) \times U(1)_Q \times U(1)_A \times U(1)_R$ symmetry in 
the $\NF=2$, $\NA=1$ model is broken down to $Sp(4) \times U(1)_{Q-A} 
\times U(1)_R$ at the point $M_{ij}=N_{ij}=O_2=O_3=0$, $P_{ij} = 
(\Lambda^4_{(2,1)} / \sqrt{2} ) \> 1_{2 \times 2} \times i \sigma_2$ on the 
quantum moduli space.  One can straightforwardly check that parton and hadron 
level calculations of the $Sp(4)^2 U(1)_{Q-A}$, $Sp(4)^2 U(1)_R$, 
$U(1)_{Q-A}^2 U(1)_R$ and $U(1)_R^3$ anomalies agree.   The singularity 
structure of the $Sp(6)$ superpotentials provides another test of 
their validity.  Branch points and poles generally signal points of enhanced 
gauge symmetry in the moduli space \cite{SeibergI}.  For example, $Sp(6)$ 
breaks to $Sp(4) \times Sp(2)$ in the $\NF=0$, $\NA=1$ effective theory when 
the antisymmetric field develops the expectation value
\be
\label{Avev2}
{\langle A \rangle = v \pmatrix{1 & & \cr
				       & 1 & \cr
				       & & -2 \cr} \times i \sigma_2.}
\ee
Expanding about this point in moduli space, we find that the nonvanishing 
$Sp(6)$ superpotential behaves as 
\be
\label{singularity}
W_{(\NF=0,\NA=1)} \sim \Bigl[ {\Lambda^{10}_{(0,1)}/ v^2 \over \Otwofour} 
\Bigr]^{\half} 
\ee
where $\Otwofour = \Tr\bigl[ (A_4 J_4)^2 \bigr]$ denotes the canonically 
normalized $O_2$ operator in the pure antisymmetric $Sp(4)$ theory.  The 
inverse squareroot singularity in the $\NF=0$, $\NA=1$ superpotential in 
(\ref{spfourflows}) is therefore recovered from (\ref{spsixflows}) 

	We could continue to apply our superpotential algorithm to theories 
with larger symplectic gauge groups.  While it becomes algebraically more 
difficult to implement as $\NC$ increases, the method does provide a general 
means for mapping the ground state structure of any model of type 
(\ref{symgroup}).  But we will instead turn to consider the implications of 
our $Sp(4)$ and $Sp(6)$ findings for theories built out of products of these 
groups.  We take up this topic in the next section.

\section{Product group models}

	The moduli spaces of supersymmetric theories based upon product gauge 
groups $G = \prod G_i$ are generally more complicated than those for simple 
group models. The nonperturbative superpotentials that summarize their 
structure involve the strong interaction scales for each factor in $G$.  
The dependence of $\Wdyn$ upon $\Lambdabi$ is not fixed by $U(1)_R$ invariance 
since scale ratios with zero R-charge can arise.  Symmetry principles also do 
not completely constrain the matter fields in $\Wdyn$ even when they all 
transform according to only fundamental or singlet irreps of $G_i$.  But by 
considering different limits in which the product group theory reduces to 
some known simple group model, one can often reconstruct its full 
superpotential.  This approach was used in refs.~\cite{ILS} and \cite{PST} to 
study the vacua of several instructive $Sp(2) \times Sp(2)$ and 
$Sp(4) \times Sp(2)$ theories.  Following these works, we will broaden the 
scope of known symplectic product group models to include a number of 
interesting $Sp(4) \times Sp(4)$ and $Sp(6) \times Sp(2)$ theories.

	The first model we investigate has symmetry group 
\be
\label{spfourxspfoursymgroup}
{G = \bigl[Sp(4)_L \times Sp(4)_R 
\bigr]_{\rm local} \times \bigl[ SU(4)_L \times SU(2)_R \times 
U(1)_Q \times U(1)_L \times U(1)_R \times U(1)'_R \bigr]_{\rm global}}
\ee
and matter content
\begin{eqnarray}
\label{spfourxspfourmattercontent}
Q_{\alpha \adot} & \sim & (4,4; 1,1; 1,0,0,0) \nonumber \\
L_{\alpha i} & \sim & (4,1; 4,1 ; 0,1,0,0) \nonumber \\
R_{\adot I} & \sim & (1,4; 1,2; 0,0,1,0) \nonumber \\
\Lambda_L^5 & \sim & (1,1; 1,1; 4,4,0,-2) \nonumber \\
\Lambda_R^6 & \sim & (1,1; 1,1; 4,0,2,0). 
\end{eqnarray}
We add a prime onto the $U(1)$ in (\ref{spfourxspfoursymgroup}) that counts 
R-charge to distinguish it from the abelian factor which tallies $Sp(4)_R$ 
flavor number.  We also adopt the nomenclature of ref.~\cite{PST} and let 
$\alpha,\beta$ and $\adot, \bdot$ respectively denote $Sp(4)_L$ and $Sp(4)_R$ 
color indices.  We use $i,j$ and $I,J$ as $SU(4)_L$ and $SU(2)_R$ flavor 
indices.  

	The moduli space of the chiral $Sp(4)_L \times Sp(4)_R$ gauge 
theory is most conveniently analyzed in the $\Lambda_L \gg \Lambda_R$ limit.  
Its dynamics in the intermediate energy range $\Lambda_R < \mu < \Lambda_L$ 
is then described by an effective $Sp(4)_L$ field theory in which $Sp(4)_R$ 
plays the role of a weakly coupled external gauge group.  The strong 
left-handed force confines the $Q_{\alpha \adot}$ and $L_{\alpha i}$ quarks 
into the mesons appearing inside the matrix 
\be
\label{Meightxeight}
{M_{8 \times 8} = 
\pmatrix{Q^T J Q & Q^T J L \cr
	 L^T J Q & L^T J L \cr},}
\ee
while the right handed fields remain unbound.  The $Sp(4)_L$ color-singlet 
hadrons 
\begin{equation}\label{intermediatematter}
\begin{array}{rclrcl}
O_1 &=& \Tr(QJQ^T J) \qquad &
(V_L)_{\adot i} &=& (Q^T J L)_{\adot i} \\
A_{\adot \bdot} &=& 
  \bigl[ Q^T J Q + \frac{1}{4} O_1 J \bigr]_{\adot \bdot} \qquad &
(M_L)_{ij} &=& (L^T J L)_{ij} 
\end{array}
\end{equation}
along with the $R_{\adot I}$ quarks thus represent the active matter degrees 
of freedom in the energy interval between the two $Sp(4)$ scales.

	The quantum numbers of the partons in (\ref{spfourxspfourmattercontent})
were intentionally chosen so that the $Sp(4)_L$ gauge group would act upon 
$\NFL=\NCL~+~2=4$ flavors of fundamental quartets.  This theory generates the 
superpotential $W_L = - \Pf M_{8 \times 8} / \Lambda_L^5$ which is analogous 
to $W_{\NF=\NC+1}$ in eqn.~(\ref{NFequalsNCplusone}) for SUSY QCD 
\cite{IntriligatorPouliot}.  After decomposing the Pfaffian of the $8 \times 8$ 
matrix in terms of its $4 \times 4$ blocks, we obtain the superpotential
\be
\label{WL}
{W_L = -{1 \over 16 \Lambda_L^5} 
\Bigr[ \bigl( O_1^2 + 16 \Pf A \bigr) \Pf  M_L + 16 \Pf(V_L^T J V_L) 
+ \e^{ijkl} (M_L)_{ij} \bigl( O_1 V_L^T J V_L 
 - 4 V_L^T J A J V_L \bigr)_{kl} \Bigr]}
\ee
for the intermediate effective theory.

	At energies below the $\Lambda_R$ scale, the $Sp(4)_R$ force among the 
three flavors of right handed quartets in $(V_L)_{\adot i}$ and $R_{\adot I}$ 
and the antisymmetric field $A_{\adot \bdot}$ grows strong.  As we know from 
our results for $\NF=3,\NA=1$ $Sp(4)$ theory, the nonperturbative $Sp(4)_R$ 
dynamics confines the right handed partons inside the colorless mesons
\be
\label{MNsixxsix}
{M_{6 \times 6} = 
\pmatrix{V_L^T J V_L & V_L^T J R \cr
	 R^T J V_L & R^T J R \cr} 
\qquad{\rm and}\qquad
N_{6 \times 6} = 
\pmatrix{V_L^T J A J V_L & V_L^T J A J R \cr
	   R^T J A J V_L & R^T J A J R \cr}}
\ee
and produces the superpotential
\be
\label{WR}
{W_R = - {\e^{ijklmn} \over 48 \Lambda_L^5 \Lambda_R^6} \bigl[ 
M_{ij} M_{kl} M_{mn} O_2 + 12 N_{ij} N_{kl} M_{mn} \bigr].}
\ee
When the $6\times 6$ matrices are decomposed in terms of their $4\times 4$, 
$4 \times 2$ and $2 \times 2$ block components, the following set of 
$Sp(4)_L \times Sp(4)_R$ invariant operators naturally emerges:
\footnote{$\>$ We add a prime on $N_L'$ as a reminder that it differs from the 
canonical $N$ meson in eqn.~(\ref{mesonops}).}
$$ 
\begin{array}{rclrcl}
O_1 &=& \Tr(QJQ^T J)  \qquad&  O_2 &=& \Tr(AJ)^2 \\
(M_L)_{ij} &=& (L^T J L)_{ij}  \qquad&  (M_R)_{IJ} &=& (R^T J R)_{IJ} \\
(N_L')_{ij} &=& (L^T J Q J Q^T J L)_{ij} = -(V_L^T J V_L)_{ij}  \qquad&
(N_R)_{IJ} &=& (R^T J A J R)_{IJ} 
\end{array} 
$$
\begin{eqnarray} \label{spfourxspfourops}
(P_L)_{ij} &=& (L^T J Q J A J Q^T J L)_{ij} 
	= -(V_L^T J A J V_L)_{ij}  \nonumber \\*
S_{iI} &=& (L^T J Q J R)_{iI} = -(V_L^T J R)_{iI} \nonumber\\*
T_{iI} &=& (L^T J Q J A J R)_{iI} = -(V_L^T J A J R)_{iI}. 
\end{eqnarray}
These operators serve as moduli space coordinates for the theory with 
$\NFL=2$ and $N_{FR}=1$ flavors of $L_\alpha$ and $R_{\adot}$ quarks and 
$N_Q=1$ $Q_{\alpha\adot}$ field.  

	The total superpotential that characterizes the far infrared 
structure of this $Sp(4)_L \times Sp(4)_R$ model simply equals the sum of 
the left and right handed sector components \cite{PST}:
\be
\label{Wonetwoone}
{W_{(N_Q=1,\NFL=2,N_{FR}=1)} = W_L + \beta W_R.}
\ee
When expressed as functions of the operators in (\ref{spfourxspfourops}), the 
two terms on the RHS become 
\begin{eqnarray}
\label{WLR}
W_L &=& -{1 \over 16 \Lambda_L^5} \Bigl[ (O_1^2 - 4 O_2) \Pf M_L
 + 16 \Pf N_L' + \e^{ijkl} (M_L)_{ij} (4 P_L - O_1 N_L')_{kl} \Bigr] \nonumber \\
W_R &=& -{1 \over 16 \Lambda_L^5 \Lambda_R^6} \Bigl\{ 4 O_2 \Pf N_L' \Pf M_R
 + 16 \Pf P_L \Pf M_R + 4 \Pf N_R \e^{ijkl} (N_L')_{ij} (P_L)_{kl} \nonumber \\
& & \qquad\qquad\qquad + \e^{ijkl} \e^{IJ} 
 \bigl[ O_2 S_{iI} S_{jJ} (N_L')_{kl}
 + 4 T_{iI} T_{jJ} (N_L')_{kl} + 8 S_{iI} T_{jJ} (P_L)_{kl} \bigr] 
 \Bigr\}.
\end{eqnarray}
The relative coefficient $\beta$ standing between these terms is convention 
dependent and {\it a priori} unknown.  But as we shall shortly see, its value 
$\beta=1/4$ is fixed by threshold factor independent matching relations and 
parity considerations in the $N_Q=\NFL=N_{FR}=1$ theory.  We should also note 
that although the superpotential in (\ref{Wonetwoone}) was derived in the 
$\Lambda_L \gg \Lambda_R$ limit, its validity transcends this special case. 
$W_{(N_Q=1, \NFL=2, N_{FR}=1)}$ describes the entire low energy moduli space 
for arbitrary values of the $Sp(4)_L$ and $Sp(4)_R$ scales \cite{PST}.

	If we couple sources to the dynamically generated superpotential in 
(\ref{Wonetwoone}) and remove heavy degrees of freedom, the $N_Q=1$, $\NFL=2$, 
$N_{FR}=1$ model flows down to theories with fewer matter fields.  We first add 
a tree level mass term which decouples one flavor of left handed quarks and 
yields the $N_Q=\NFL=N_{FR}=1$ model whose moduli space is defined by three 
quantum constraints:
\begin{mathletters}
\label{spfoursqconstraints}
\begin{eqnarray}
(O_1^2 - 4 O_2) \Pf M_L - 4 O_1 \Pf N_L' + 16 \Pf P_L &=& 16 \Lambda_L^6 
  \label{sp4xsp4cons-a} \\*
\Pf N_L' \Pf N_R + \Pf M_R \Pf P_L + \e^{ij} \e^{IJ} S_{iI} T_{jJ} 
 &=& - \Pf M_L \Lambda_R^6 \label{sp4xsp4cons-b} \\*
O_2 (\Pf N_L' \Pf M_R + \det S) + 4 (\Pf P_L \Pf N_R + \det T) 
 &=& (O_1 \Pf M_L - 4 \Pf N_L') \Lambda_R^6. \label{sp4xsp4cons-c}
\end{eqnarray}
\end{mathletters}
The $\Lambda_L$ scale in (\ref{sp4xsp4cons-a}) is related to its 
counterpart in (\ref{WLR}) by a simple matching condition, while the 
$\Lambda_R$ scales in the upstairs and downstairs theories are exactly the 
same.  All mesons appearing in these formulae represent $2 \times 2$ 
matrices in flavor space.  

	The three quantum relations in (\ref{spfoursqconstraints}) do not show
any sign of a discrete parity reflection in the $N_Q=\NFL=N_{FR}=1$ model 
whose left and right handed sectors are precisely parallel.  But appearances 
can be deceiving.  After eliminating the $P_L$ field via the first constraint 
and performing the superfield redefinitions
\begin{eqnarray}
N_R' &=& R^T J Q^T J Q J R = N_R + \frac{1}{4} O_1 M_R \nonumber \\*
\Delta &=& \det Q = \Pf Q^T J Q = {1 \over 16} (O_1^2 - 4 O_2), 
\end{eqnarray}
we find two new quantum relations which we incorporate within the 
superpotential
\begin{eqnarray}
\label{Wonetwotwo}
&&  W_{(N_Q=\NFL=N_{FR}=1)} =
X \Bigl[\Delta \Pf M_L \Pf M_R - \Pf N_L' \Pf N_R' 
  - \e^{ij} \e^{IJ} S_{iI} T_{jJ} - \Pf M_R \Lambda_L^6 - \Pf M_L \Lambda_R^6 
\Bigr] \nonumber \\
&& \qquad\qquad + Y \Bigl[ \Delta ( \Pf M_L \Pf N_R' + \Pf M_R \Pf N_L') 
 - {1 \over 2} O_1 \Pf N_L' \Pf N_R' + (\Delta - {1 \over 16} O_1^2) \det S 
 - \det T \nonumber  \\
&& \qquad\qquad\qquad
 - {1 \over 4} O_1 \e^{ij} \e^{IJ} S_{iI} T_{jJ} 
 - \Pf N_R' \Lambda_L^6 - \Pf N_L' \Lambda_R^6 \Bigr].  
\end{eqnarray}
{\it Mirabile dictu,} this expression is manifestly left-right symmetric!  

	If we continue to add quark mass terms, we can flow down to the 
$Sp(4)_L \times Sp(4)_R$ model that has only $Q_{\alpha\adot}$ matter.  Since 
no source terms in the $ N_Q=\NFL=N_{FR}~=~1$ theory transform like $(2,2)$ 
under the global $SU(2)_L \times SU(2)_R$ chiral symmetry group, the $S$ and 
$T$ meson fields cannot develop nonzero expectation values.  When we evaluate 
the equations of motion for the remaining $M_{L,R}$ and $N'_{L,R}$ mesons 
along with the $X$ and $Y$ Lagrange multipliers, we find two distinct branches 
for the dynamical superpotential in the $N_Q=1$, $\NFL=N_{FR}=0$ theory:
\be
\label{Wonezerozero}
W_{(N_Q=1,\NFL=N_{FR}=0)} = \cases{
\displaystyle{\bigl[ \Lambda_L^{7 \over 2} \pm  \Lambda_R^{7 \over 2} 
 \bigr]^2 \over \Delta} \cr
\cr
\displaystyle{1 \over \Delta} \Bigl[ \Lambda_L^7 + \Lambda_R^7 + 
\displaystyle{{O_1 \over \sqrt{O_1^2-16 \Delta}} \Lambda_L^{7 \over 2} 
\Lambda_R^{7 \over 2} \Bigr].}}
\ee
This result exhibits several interesting features.  Firstly, it correctly 
reduces to the instanton generated $\NF=2$ $Sp(4)$ superpotential in 
(\ref{Wquark}) when either $\Lambda_L$ or $\Lambda_R$ vanishes.  Secondly, its 
dependence upon the dimensionless and chargeless ratio $R=16 \Delta / O_1^2$ 
could not have been determined by symmetry considerations or integrating in 
techniques.  We also note that the full superpotential cannot be obtained by 
considering the $\Lambda_L \gg \Lambda_R$ limit of the pure $Q_{\alpha \adot}$ 
matter theory.  Instead, it is necessary to start from a theory with a 
genuine moduli space of vacua and integrate out matter to obtain 
eqn.~(\ref{Wonezerozero}).  Finally, the dichotomy in 
$W_{(N_Q=1, \NFL=N_{FR}=0)}$ directly reflects the two $\NF=0, \NA=1$ $Sp(4)$ 
superpotential branches in eqn.~(\ref{spfourflows}).  When 
$Sp(4)_L \times Sp(4)_R$ is broken to its diagonal subgroup by the expectation 
value $\langle Q \rangle = v \> 1_{4 \times 4}$, the fluctuations about 
this vev which survive in the low energy $Sp(4)_{L+R}$ theory transform 
according to the 5 dimensional antisymmetric representation.  Expanding about 
this point, we find the operator relations
\begin{eqnarray}
\label{oprelns}
O_1 &=& -4v^2 - O_2^{(4)} \nonumber \\
\Delta &=& \det Q = v^4 + O( v^2 O_2^{(4)}) 
\end{eqnarray}
and matching condition $\Lambda_L^{7 \over 2} \Lambda_R^{7 \over 2} = v^3 
\Lambda_{L+R}^4$.  When these expressions are inserted into the terms inside 
the curly brackets in (\ref{Wonezerozero}), both $Sp(4)$ 
$W_{(\NF=0, \NA=1)}$ superpotentials are recovered as limits of 
$W_{(N_Q=1, \NFL=N_{FR}=0)}$ in the $Sp(4)_L \times Sp(4)_R$ theory.

	Techniques similar to those which we have used to analyze these $Sp(4) 
\times Sp(4)$ models can be applied to other product group theories as 
well.  For example, we briefly sketch the outline of a model based upon 
$Sp(6) \times Sp(2)$ with $(6,2)$ matter that incorporates our simple 
$Sp(6)$ group findings.  It is again easiest to first consider the 
limit $\Lambda_2 \gg \Lambda_6$.  At energies well above the $\Lambda_6$ 
scale, the $Sp(2)$ gauge dynamics confine the elementary fields into $Sp(2)$ 
singlet mesons $A_{\alpha \beta}$ and generate the superpotential 
$W_2 = -\Pf A/\Lambda_2^3$.  For $\mu < \Lambda_6$, the strong $Sp(6)$ force 
binds together the composite antisymmetric fields into completely colorless 
$O_2$ and $O_3$ combinations and produces a dynamic superpotential $W_6$ 
which can be read off from eqn.~(\ref{spsixflows}).  The total superpotential 
$W = W_2 + \gamma W_6$ for the low energy $Sp(6) \times Sp(2)$ effective 
theory equals the sum of the two separate contributions with a relative 
coefficient $\gamma$ that depends upon the normalization conventions for 
$\Lambda_2$ and $\Lambda_6$.  Variations on this model with additional $Sp(6)$ 
quark fields can be worked out along similar lines. 

\section{Conclusion}

	In this article, we have examined the confining phase vacua of 
several symplectic SUSY gauge theories with matter in fundamental and 
antisymmetric representations.  These models exhibit interesting 
nonperturbative features such as multiple quantum constraints, intricate 
superpotentials depending upon chargeless operator ratios and singularity 
structures that reproduce underlying theories at points of enhanced gauge 
symmetry.   Our approach to studying these particular $Sp(2\NC)$ 
models can be profitably applied to other confining supersymmetric theories. 
In simple gauge group models, it is often possible to adjust microscopic 
matter contents so that the dynamically generated superpotentials are 
proportional to polynomials in gauge invariant fields.  These polynomials are 
determined by requiring that they yield hadronic equations of motion which 
reduce to classical constraints among parton constituents.  When tree level 
masses are added and heavy fields are integrated out, nonperturbative 
superpotentials for models with fewer matter degrees of freedom can be 
derived.  This top-down algorithm yields complete towers of low energy 
effective theories.  In contrast, the bottom-up approach which starts with pure 
glue theory and successively integrates in matter frequently fails at various 
rungs on the effective theory ladder.  While the utility of integrating in 
techniques has been demonstrated in certain cases \cite{ILS,IntriligatorII}, 
our experience with the symplectic models in this paper leads us to believe 
the top-down approach is more generally useful.

	Once the ground state structure of a confining simple group model is 
known, its impact upon product group theories is largely determined.  By 
considering various limits in which a product group model reduces to a known 
simple group theory, one can reconstruct its full superpotential.  Simple 
group models thus serve as building blocks for arbitrarily complicated 
theories.

	In closing, we mention some extensions of this work which would be 
interesting to pursue.  The top-down method we have followed yielded 
exact superpotentials for $Sp(4)$ and $Sp(6)$ theories.  In principle, 
it can be used to construct the low energy effective theory for any specified 
value of $\NC$.  But deriving closed form expressions for confining phase 
superpotentials in symplectic theories with arbitrary numbers of colors would 
be preferable.  Unfortunately, we have not yet found a clear pattern among the 
special cases we have solved so far.  It would similarly be interesting to 
examine entire sequences of product group theories.  Large $\NC$ limits of 
such models might reveal unexpected surprises.  Finally, constructing dual 
descriptions of multimatter symplectic theories in nonconfining phases with 
zero tree level superpotentials remains an important outstanding problem.  We 
look forward to investigating these issues in the future.

\bigskip\bigskip\bigskip
\centerline{{\bf Acknowledgments}}
\bigskip

	We thank Sandip Trivedi for many helpful discussions.  The work of 
PC and PK was supported in part by the DuBridge Foundation and by the 
U.S. Dept. of Energy under DOE Grant no. DE-FG03-92-ER40701.


\end{document}